# Linguistically inspired roadmap for building biologically reliable protein language models

Mai Ha Vu[1,C], Rahmad Akbar[2,], Philippe A. Robert[2,], Bartlomiej Swiatczak[3,], Geir Kjetil Sandve[4,*], Victor Greiff[2,*], Dag Trygve Truslew Haug[1,*,C]

[1] Department of Linguistics and Scandinavian Studies, University of Oslo, Norway

[2] Department of Immunology, University of Oslo and Oslo University Hospital, Norway

[3] Department of History of Science and Scientific Archeology, University of Science and Technology of China, China

[4] Department of Informatics, University of Oslo, Oslo, Norway

Equal contribution

[*] Joint supervision

[C] Correspondence: m.h.vu@iln.uio.no, d.t.t.haug@ifikk.uio.no

**Abstract**

Deep neural-network-based language models (LMs) are increasingly applied to large-scale protein sequence data to predict protein function. However, being largely black-box models and thus challenging to interpret, current protein LM approaches do not contribute to a fundamental understanding of sequence-function mappings, hindering rule-based biotherapeutic drug development. We argue that guidance drawn from linguistics, a field specialized in analytical rule extraction from natural language data, can aid with building more interpretable protein LMs that are more likely to learn relevant domain-specific rules. Differences between protein sequence data and linguistic sequence data require the integration of more domain-specific knowledge in protein LMs compared to natural language LMs. Here, we provide a linguistics-based roadmap for protein LM pipeline choices with regard to training data, tokenization, token embedding, sequence embedding, and model interpretation. Incorporating linguistic ideas into protein LMs enables the development of next-generation interpretable machine-learning models with the potential of uncovering the biological mechanisms underlying sequence-function relationships.

# 1 Introduction

A growing number of studies apply machine learning tools called Language Models (LMs) or sequence models (e.g., BERT[1], RoBERTa[2] and GPT[3]) to biological sequence data[4–19]. LMs are a tool in Natural Language Processing (NLP), a subfield of computer engineering applied to natural language. They create probability distributions over sequences of tokens (e.g., characters, words, subwords) by learning statistical patterns from large, unlabeled sequence data in a self-supervised manner[1,3,20]. As such, LMs have become a popular tool for processing the exponentially growing number of unlabeled protein sequences[21–23], with the potential to capture biochemical and physicochemical principles that underlie sequence-structure-function relationships[5,12,24,25]. With the help of interpretability methods, these captured principles could be extracted to advance fundamental protein science research on the one hand, and accelerate rational protein design, including therapeutics design, on the other[12,24,26]. To harness this potential, protein LMs must have verifiably learned the true generalizable scientific principles underlying the data without overfitting for a particular training dataset. So far, it remains unclear how much biological knowledge (e.g., structural information, functional building blocks) is required for building such protein LMs.

Just as protein science is the relevant domain knowledge for protein LMs, linguistics, the field that studies natural language, is the relevant domain knowledge for natural language LMs. Given the success of high-capacity natural language LMs without linguistic priors on many NLP tasks[27–30], it might be tempting to also build protein LMs with minimal biological priors. Indeed, multiple protein LMs with minimal biological priors still perform remarkably well on structure and function prediction tasks[4,5,10,12,16,25,31]. However, there is no guarantee that the patterns learned by these LMs are biologically explanatory rules. In fact, an in-depth comparison of protein sequences to natural language suggests that greater integration of biological domain knowledge in protein LMs is needed precisely due to the *differences* between the two types of data (Figure 1).

The success of natural language LMs without linguistic priors is primarily due to the fact that linguistic data conforms to the distributional semantics hypothesis: tokens that share similar context have similar semantic meaning[32]. Consequently, some semantic information can be inferred in a self-supervised manner from distributional properties alone, if the input tokens correspond to individual meaning bearing units. Because many natural language orthographies contain built-in symbols (e.g., space, punctuation) that divide linguistic sequences into meaningful tokens, explicit linguistic knowledge is often unnecessary to obtain an adequate generalization of the linguistic data. It is so far unknown whether distributional semantics is applicable to proteins, because there is no comprehensive definition of meaningful protein tokens (i.e., 'protein words'). Since protein sequences do not have built-in symbols to indicate structure, a more analytical, expertise-driven approach to finding 'protein words' is needed.

Furthermore, biological data and linguistic data differ in their coverage of all possible *types* of sequences. Currently, there is no comprehensive knowledge of protein sequence-function mapping rules, and thus there is no guarantee that the available data contains all relevant information. For more specialized language models, such as immune receptor language models, the available sequence data undersamples the potential sequence space ($10^9$ publicly available immune receptor sequences[22,23] vs. >$10^{14}$ biologically possible sequences[33,34]). In comparison, due to pre-existing linguistic knowledge, natural language corpora of well-studied languages are verifiable for comprehensiveness. In the case of under-resourced languages, models especially benefit from linguistic priors compared to languages that have an abundance of data available[35–37]. Similarly, under-resourced protein families might also benefit from biological priors incorporated into protein LMs.

Lastly, deep protein LMs are typically trained on protein amino acid sequences without their original genomic context[12]. In the absence of genomic context, which would provide additional clues to protein function due to the clustering of functionally related genes[38], there is no larger context from which the meaning (i.e., function) of individual sequences could be inferred in protein LMs. In contrast, in natural language corpora, the larger context of other sentences can aid significantly in inferring the meaning of a target sentence. Thus protein LMs must rely more on domain knowledge to determine the overall function of isolated sequences[39].

While our points echo previous discussions on the differences between natural language and protein sequences[4,5,12,24,25], our approach is unique in the focus on building protein LMs that learn true biological principles and in drawing insights from linguistics to reach this goal. In particular, the recent benchmarking study by Unsal et al.[12] also called for incorporating more protein knowledge into the models. However, unlike outlined here, they did not draw from previous studies of natural language LMs, and their goal was to improve protein LM performance rather than to build biologically reliable protein LMs.

In this Perspective, we examine multiple aspects of the deep LM pipeline[27,40,41]: specifically, pre-training data selection, tokenization, token and sequence embedding, and interpretability methods (Figure 2) in protein LMs, drawing from studies of natural language LMs. Our suggestions are meant to pave the way for future exploratory research directions that would systematically address current LM-related challenges.

## 2 Pre-training data should reflect goals of downstream task

During pre-training, an LM generates a probabilistic model of large sequence data through self-supervision, typically by identifying missing tokens in a text. The set of sequences that the LM aims to model is the *language.* Because it is impossible to exhaustively list all sequences of an unboundedly large language, only a sample is provided to the model in the pre-training dataset. Data points in the pre-training dataset thus define the language that LMs model: for example, BERT, an LM pre-trained on English Wikipedia and an English book corpus would be a model of formal English[1], while BioBERT, which is pre-trained on biomedical texts[42] is a model of biomedical English. In this sense, "language" does not necessarily align with the conventional definition of a natural language, such as Norwegian, Indonesian, or Swahili.

The language of protein LMs is thus determined by the set of protein sequences that are included in the pre-training data: it could be all available protein sequences for a broad protein language model, the set of human protein sequences for a human protein language model, or the set of all observed antibody sequences for an antibody language model. In any case, it is imperative to have a rigorous definition of the language informed by the scientific goals of the protein model, and to construct a pre-training dataset that reflects this language with suitable information for inferring generalizable biological principles.

While several studies analyze the effects of pre-training data choice on natural language LM behavior[43–46], for protein LMs most studies benchmark different datasets for model performance, without a similarly in-depth discussion about the results[8,13].

One option is to create a language model specific to a narrow downstream task. For example, several subject-specific natural language LMs have been developed to capture subject-specific token meaning (e.g., BioBERT for biomedical texts[42]). An analogous protein LM is one that is specific to a particular type of protein, such as the antibody-specific LMs AntiBERTa[48], AntiBERTy[49], and AbLang[50].

An alternative is to use the most general and largest possible dataset for pre-training, which can be leveraged in multiple different downstream tasks. For example, multilingual LMs are pre-trained on data from multiple natural languages at once, and are fine-tuned on tasks that only pertain to one of the languages. Multilingual LMs can be particularly useful in cases where labeled data for the target language is very limited, but there is an abundance of labeled data for another language.

General protein LMs that are trained on all available protein sequences are similar to monolingual natural language LMs if the downstream task pertains to general protein properties (e.g., secondary structure, amino acid contact in the structure, and stability[10]). They are more similar to multilingual LMs if the downstream task is relevant to only a small subset of proteins (e.g., antibody sequence questions, such as antibody affinity maturation or epitope prediction).

Studies in NLP demonstrated that multilingual LMs remain limited compared to monolingual LMs; their performance correlates with the size of training data[51–54], and with the similarity between the languages in the training and testing data[55]. Thus LMs pre-trained on all available protein sequence data are likely to be most effective for downstream tasks that predict general features of proteins. To answer questions that are specific to only certain types of proteins, such as antibody sequences, specialized antibody LMs perform better[22,48,49,56–58]: for example, AntiBERTa outperforms ProtBERT, a general protein model[8] on a number of antibody-specific questions[48].

As LMs have shown the capability for knowledge transfer even between dramatically different types of data[59,60], such as linguistic and protein data, good LM performance alone cannot justify pre-training data selection for the purposes of discovering biological principles, and a careful study of the problem and sequence distribution remain necessary.

In order to choose the appropriate pre-training data that can contribute to true scientific insights, there thus needs to be careful consideration of whether it contains information transferrable to the downstream task, more empirical study to determine the viability of different types of pre-training data for various fine-tuning tasks compared to a baseline randomly generated data, and more available large datasets for specialized types of proteins. The last two points may be addressed with computational simulations[61–64], which can generate arbitrarily large datasets with a priori defined rules to test different approaches.

## 3 Tokenization should aim for biologically meaningful units

Tokenization is the subdivision of input sequences into discrete units; it is not sequence encoding, which concerns the representation and information included in the sequence (e.g., amino acid content, structural information, information about physicochemical properties, or other functions[66]). Tokenization is a fundamental step in an LM pipeline because the pre-training task typically involves the prediction of tokens in a sequence (Figure 2). We argue that beyond technical needs, finding tokens that approximate biologically meaningful motifs (similar to *linguistically* meaningful tokens in natural language) is integral to building models with biologically generalizable rules in protein LMs.

Tokenization in NLP serves both computational and ideally, linguistic goals. Computationally, tokenization helps reduce data sparsity: it enables the representation of unseen sequences as a combination of already seen tokens drawn from a finite vocabulary[41]. Thus to fulfill the computational goals, the vocabulary must be finite but exhaustive to avoid out-of-vocabulary tokens. Furthermore, tokenization is preferably automated at a large scale and results in an LM with low information entropy distribution for a given vocabulary size, ensuring lower perplexity (the ability for a model to predict a sample)[67,68] (Figure 3). Linguistically, tokens should correspond to atomic units carrying *meaning* that cannot be inferred from the characters alone. Such meaningful tokens have been traditionally derived through linguistic analysis.

In current NLP practice, tokenization methods trade-off between the computational and linguistic requirements. They can range from simple heuristics such as space-delimited tokenization to information-theoretic methods such as Byte-Pair-Encoding (BPE)[69], to hand-crafted tokens derived from linguistic analysis[41]. New tokenization algorithms are in active development within NLP, and studies have found that the effectiveness of different tokenization methods depends on the intended task, the language, and available data[35,41,68,70,71]. Additionally, because in many natural languages, simple tokenization methods, such as space-delimited tokenization, already generate tokens that overlap with linguistically desirable tokens, these methods can adequately fulfill both computational and linguistic goals. Protein sequences do not have similarly clear delimiters for establishing meaningful units.

Current protein LMs overwhelmingly use simple tokenization heuristics[6–9,31,72,73] and to a lesser extent, information-theoretic ones, such as BPE[7,15,19,74–76]. The performance of information-theoretic tokenization on various structural and function tasks remains mixed compared to simpler tokenization methodologies[7,15,76].

Both simple and information-theoretic tokenization methods generate a finite vocabulary, and information-theoretic ones, by definition, also generate models with relatively low entropy (Figure 3). Older studies have estimated entropy rate with amino acid-based tokenization to be 2.4-2.6[77], while the estimated entropy rate for English with character-based tokenization is 0.6-1.75[67,78]. These types of information-theoretic measures are not consistently reported for new protein LMs, and thus are hard to compare to existing natural language LMs. Significant divergence in entropy between protein and natural language LMs would indicate a possibility for better alternatives.

As for generating meaningful tokens, current protein tokenization methods remain inconclusive. While many studies have shown that these methods can generate token embeddings that cluster along physicochemical properties of individual amino acids[6,58,79], physicochemical properties do not map to global protein functions in a straightforward and human-interpretable way. In other studies, these tokens were also compared for similarity to experimentally verified motifs[75], defined as being overrepresented substrings in protein sequences[75], while their specific functional meaning remains unknown. Furthermore, studies in NLP have shown that information-theoretic algorithms, though widely used due to their efficiency, are not reliable for finding linguistically sound tokens[71,80]. In the absence of well-defined, biologically meaningful protein tokens, a truly informative evaluation for the linguistic criterion remains impossible.

An underexplored possibility is building a rule-based tokenizer grounded in biological domain knowledge (Figure 3). This method might especially benefit modeling certain types of proteins, where sequence data is less abundant, similarly to how linguistically guided tokenization leads to better results for under-resourced languages[35–37,81]. It is yet unknown whether these types of tokens would fulfill the computational criteria; in natural language, entropy level has been shown to be relatively low for linguistically defined tokens[65], and meaningful protein tokens might yield similar results. Still, as there is very limited knowledge of what should constitute a meaningful protein token, a rule-based tokenizer remains extremely challenging to implement. As a first step, protein scientists must define possible functional meanings that can realistically be linked to discrete protein tokens.

One possibility is to draw from existing studies that found functionally and structurally significant, subdomain-sized tokens[82–86]. The main challenge is that many expert-defined tokens require structural data, and cannot be applied to unlabeled sequence data at scale. Another possibility is to train a tokenizer based on a large number of defined tokens in protein simulations; for example, simulated antibodies in high resolution can give information about interacting and non-interacting segments[61], which can then be used to train a tokenizer. A third possibility is to train a tokenizer based on the fine-tuning task[87]. In the last two cases, further evaluation would be necessary to experimentally validate the tokens as desired ground truth in real-world data.

Biologically meaningful protein tokens might be discontinuous, overlapping, and each sequence might need to map to several different tokenization possibilities (Figure 3)[61,86], requiring alternative LM technology to appropriately process. One naive solution would be to shift from non-continuous tokens to smaller continuous subtokens with learned long-distance dependency between them, but this strategy falls short of the linguistic goal, and conceptually mixes tokenization with long-distance dependency rules at the expense of interpretability. Another possibility employed in NLP, is to re-order non-continuous tokens[88,89], so they become continuous.

Altogether, there needs to be a definition of biologically sound protein tokens that can serve as ground truth for comparing various tokenization methods. The fact that proteins contain units that compositionally determine their function (at least at the scale of protein domains[90]), similarly to how linguistic tokens compositionally map to sentence meaning, suggests that analytic, linguistic tokenization methods may be transferable to protein tokenization, given more robust data and investigations.

## 4 Token embeddings should capture protein function

LMs represent tokens as multidimensional vectors called embeddings that ideally reflect the functional meaning of these units. Token embeddings are initially calculated during pre-training, and then are further refined during fine-tuning. Embeddings can be extracted from the hidden layers of the pre-trained LM and leveraged as input for downstream tasks that use much smaller datasets (Figure 2)[1]. The linguistic function of a token embedding is to reflect the linguistic role of the token in the text. In the case of protein tokens, these roles are equivalent to their biological functions.

By pre-training LMs to predict tokens based on their context in linguistic data, the token embedding reflects the token's context, which presumably correlates with its lexical meaning according to the distributional semantics hypothesis[32]. The information captured by the embedding is typically arithmetically validated (e.g., close clustering of synonymous tokens, calculations such as king-man+woman≈queen[91])–. All validation methods require a priori knowledge about token meaning. Furthermore, there are several ways for two words to be related (e.g., king-queen and king-chief can both be closely related pairs), so similarity is limited to only certain attributes[92].

Most current protein LMs directly borrow standard NLP pre-training tasks for token embedding, without further justification[6,8,9,48,58,79,93]. The tokens, which often are amino acids, are usually represented with one-hot encoding, and the context is the rest of the protein sequence itself. Protein LMs have closely followed the evolution of embedding methods in NLP, moving from non-contextual token embedding techniques such as word2vec[58,75,79,94] where each token has a fixed embedding vector, to contextual token embeddings such as ELMo[8,9,31,72,95] and BERT[1,8,10,48], where the token embeddings depend on context. As biological function is typically encoded in several non-linear, long-distance dependencies[86], models that yield contextual word embeddings are likely more appropriate for proteins.

In alternatives to standard token embedding methods, the pre-training task is more specific to downstream tasks. For example, ProteinBERT[7] was pre-trained on protein sequences encoded together with their Gene Ontology (GO) annotation, which is an annotation of the protein sequence function. As a result, the embeddings of single amino acids captured information of both the sequence and its GO annotation. Another possibility is to pre-train the protein LM on protein-specific tasks, in addition to the self-supervised language modeling task, for example on structural information prediction[4,96].

The disadvantages of using biological information in the pre-training task are the lack of available data and potential data leakage. These tasks rely on available large-scale data that is annotated with biological information (e.g., GO information, structural information), which remains limited for certain types of proteins, such as antibody sequences[26]. Secondly, data leakage is the accidental inclusion of information in the training phase about the test data, and it can cause overly optimistic results[12,97]. Unsurprisingly, protein LMs pre-trained on tasks that were related to the fine-tuning task performed better[12]. More careful systematic studies need to accompany non-standard pre-training, and they should be tested on tasks that are unrelated to the pre-training task to ensure that the LM has learned generalizable principles.

Token embeddings extracted from protein LMs have been validated through clustering plots[6,9,58,79] and through performance on downstream tasks[4,7]. All studies clustered token embeddings in terms of physicochemical properties[6,9,58,79], as they are well-defined, and none examined whether the tokens can capture more abstract functional meaning, which would have been illuminating.

For sequence-based prediction tasks, token embeddings are the sole source for deriving sequence embeddings, since protein sequences are not trained on their broader context. Currently the most popular method for calculating protein sequence embedding is through average pooling (i.e., the average of the token embeddings)[6,8–10,98]. In contrast, linguistics provides a principled, rule-driven method for deriving sentence meaning from structure compositionally[99], and there are multiple structure-sensitive sequence embedding techniques in NLP[100,101]. For protein LMs, it remains an understudied question whether structure-sensitive sequence embeddings could prove viable[98].

In summary, protein LMs lack an extensive investigation into the information contained by token embeddings, as all evaluation remains superficial with plotting physicochemical properties and indirect with benchmarking on downstream tasks. We argue that with domain-based tokenization and carefully chosen token embedding tasks, token embeddings could capture more abstract biological functions that go beyond physicochemical properties. Such token embeddings would also significantly improve the interpretability of the protein LMs.

## 5 Type of interpretability method affects learnable patterns

Although deep LMs have proven a powerful tool for modeling sequence data, the patterns they learn remain hidden. Recently, the need for transforming these black-box models into transparent, interpretable glass-box models has resulted in efforts to obtain better interpretability from deep natural language LMs[40]. Improving model interpretability is crucial to better understand what these models learn and to pinpoint the causes for their failures. Among others, linguistics-inspired methods to probe LMs have been popular in NLP research[105,106,108]. Similar efforts, however, have not been widely adopted for protein LMs[25], and we argue that incorporation of interpretability and explainability concerns should be an essential part of protein LM design from the start.

Besides advancing scientific knowledge, inferring rules from protein LMs with interpretability methods is an integral step in the rational protein engineering pipeline[26] (Figure 4A). A precondition for successful sequence-function rule extraction is that the protein LM has reliably learned general biological principles, which can be facilitated by incorporating relevant domain knowledge into the LM design. Various interpretability and rule-extraction methodologies can then help extract the biological principles hidden within the black-box model, which can inform rational protein design and novel protein synthesis[12,24,26]. It is crucial that the inferred rules are used as guidance rather than true answers about biology, and that they are experimentally validated. The novel proteins designed based on the inferred rules can in turn serve as additional data for further LM training after experimental validation. Rigorous interpretability-focused examination can also help evaluate whether well-performing models have learned truly meaningful representations and sequence-function mappings[114–116].

We distinguish three types of interpretability methods: architecture analysis, linguistics-inspired experimentation, and grammatical inference (Figure 4B). The choice of method biases the type of information that can be learned about the LM and the modeled sequences, so it is crucial to be aware of the limitations inherent in each method. Broadly, there are two types of information that can be gained: the localization of specific types of knowledge in the architecture, and the specific sequence-function rules that the model has successfully learned.

Architecture analysis (e.g., studying specific layers in the architecture, probing the pre-trained embeddings, interpreting attention pattern heatmaps and saliency maps), the most popular interpretability method, can yield information about where and how the model architecture stores various types of knowledge about the sequence and highlight parts of the input that were significant for the classification task. Understanding the localization and method of knowledge storage in the architecture can improve the explainability and efficiency of the model. At the same time, the patterns extracted from architecture analysis do not have a straightforward and reliable correspondence to real interaction between tokens, and thus can only be employed for rule extraction after careful validation of the model.

In natural language BERT, analysis of attention patterns revealed that different, unpredictable parts of BERT specialize in specific aspects of the syntactic structure (e.g., direct objects attend to verbs in one head, but not others)[117]. While valuable as a way for finding known relations in the model, these results also suggest that the full syntactic structure of a sentence cannot be straightforwardly extracted from attention patterns alone. The same precautions should be exercised when applying analogous methods for protein LM interpretability. Even though attention patterns in protein LMs were shown to correlate with amino acid contact in the protein structure[25,48,56], the usefulness of the method remains limited for rational protein design, as it would be difficult to know *which* attention patterns correlate with real amino acid interactions in protein sequences. Furthermore, these types of explainability methods are often reliant on subjective human interpretation[118,119] and are demonstrated on a very small number of examples[48,49]. Simulated ground truth data could help with more controlled and robust examination of architecture patterns.

The two other methods, linguistics-inspired experimentation and grammatical inference could yield generalizable, well-defined sequence-function rules that the model has learned. In linguistics-inspired experimentation, researchers test the knowledge of the model for a hypothesized sequence rule by feeding it pairs of sentences that differ minimally (ideally by one token), where one follows the rule and the other violates it[103–108]. If the LM successfully distinguishes between the two sentences, it has likely learned the rule. For example, BERT can distinguish near perfectly between sentences such as the grammatical "*The game that the guards hate is bad*" and the ungrammatical "*The game that the guards hate are bad*", indicating that BERT has learned that verbs (e.g., *is, are*) must agree in number with the subject (e.g., *the game*), even when there are intervening nouns (e.g., *the guards*)[104].

For protein LMs, the method could be applied to extract new sequence-function rules. A challenge is that it requires concrete hypothesized rules that clearly distinguish between two classes of protein sequences. Furthermore, since the goal is to learn new structure-function relationship rules, it requires a substantial amount of guesswork. For example, if protein scientists hypothesize that a certain motif is responsible for a certain protein function, then they could test the veracity of this rule by feeding the protein LM sequences that contain the motif and sequences that do not. In other words, the LM can act as a fast proxy to experimental testing.

Finally, there is a long history of developing grammatical inference algorithms with the purpose of extracting grammar (i.e., a set of rules) from a set of strings[120,121]. For example, Angluin's L* algorithm can learn a finite-state automaton that describes a set of strings, if membership and equivalence queries are allowed from the oracle[121]. Recently, Weiss et al. have used the L* algorithm to obtain a finite-state automaton from an RNN, where the RNN itself serves as the oracle[111].

For protein LMs, an example of a rule extracted this way could be a sequence pattern described as a regular expression (e.g. [DC]|[A.].*G..P) corresponding to a given biological function, from an LM that is trained to recognize only proteins that express that function. The advantage of grammatical inference compared to linguistically inspired methods is that it does not require concrete hypothesized rules to use, although it still requires a priori restriction on the *class* of possible rules, as that determines the algorithm[122]. As of now, efficient algorithms only exist as proof of concept, for relatively simple models (mainly RNNs) and types of rules, and cannot handle noisy input data well. Nevertheless, these methods could become useful as the field develops better algorithms[112,113].

In summary, well-designed and well-performing protein LMs can only reach their full potential to be useful for rational protein design as transparent glass-box models with thoroughly understood, interpretable rules. We have identified three types of methodology (architecture analysis, linguistic experimentation, and grammatical inference), and each of them has different sets of advantages and disadvantages. All of them require better a priori understanding of biological rules that can exist in protein sequences. Therefore, in order to access a transparent understanding of a protein LM, there needs to be ongoing probing of the LM that uses a diversity of methodology. The employment of ground truth (simulated) data can further help with examining these interpretability methods[61,123].

## 6 Conclusions

Similarities between protein and natural language sequences have inspired the use of LMs for protein sequences, which were originally tools for modeling linguistic sequences. Self-supervised protein LMs have the potential for identifying relevant sequence rules that can be further experimentally tested, and thereby contributing to fundamental questions in biological research and accelerating the rational protein therapeutics design. However, current practice in designing and building protein LMs have fallen short of appropriately adapting these models to protein sequences due to a lack of a deeper understanding on how they were originally built for modeling linguistic sequences. In this Perspective, we have highlighted various parts of the LM pipeline (pre-training data, tokenization, token and sequence embedding, and rule extraction), and we have shown how understanding the original linguistic intent underlying each of these steps can inform the building of more biologically informed protein LMs that answer specific downstream questions of interest.

Throughout, we called for more thorough benchmarking experiments that would compare how different design choices affect not only performance on prediction tasks, but also the types of patterns LMs learn. Benchmarking can be effectively performed on simulated data that encodes the most essential properties of the protein sequence data[61,123] to serve as proof of concept for the specific model used. Furthermore, it can be sufficient to benchmark small models that differ on only one variable (e.g., type of tokenization), as has been done for the testing of different types of encodings[66]. The goal is not to build the best-performing model, but to systematically investigate aspects of LMs that may make a difference for successful sequence-function rule learning before scaling to larger models. Protein LMs that have thoughtfully incorporated the considerations discussed at each point are then more likely to have learned the relevant biological rules for the modeled sequences, and thus may be better suited for successful rule-extraction for rational protein design.

## Funding

We acknowledge generous support by The Leona M. and Harry B. Helmsley Charitable Trust (#2019PG-T1D011, to VG), UiO World-Leading Research Community (to VG), UiO:LifeScience Convergence Environment Immunolingo (to VG, GKS, and DTTH), EU Horizon 2020 iReceptorplus (#825821) (to VG), a Research Council of Norway FRIPRO project (#300740, to VG), a Research Council of Norway IKTPLUSS project (#311341, to VG and GKS), a Norwegian Cancer Society Grant (#215817, to VG), and Stiftelsen Kristian Gerhard Jebsen (K.G. Jebsen Coeliac Disease Research Centre) (to GKS).

## Acknowledgements

We thank Kyunghyun Cho (New York University, New York, NY, USA) and Emily M. Bender (University of Washington, Seattle, WA, USA) for their very helpful comments on the manuscript.

## Declaration of interests

V.G. declares advisory board positions in aiNET GmbH, Enpicom B.V, Specifica Inc, Adaptyv Biosystems, EVQLV, and Omniscope. V.G. is a consultant for Roche/Genentech, immunai, and Proteinea. The remaining authors declare no competing interests.

# Figures

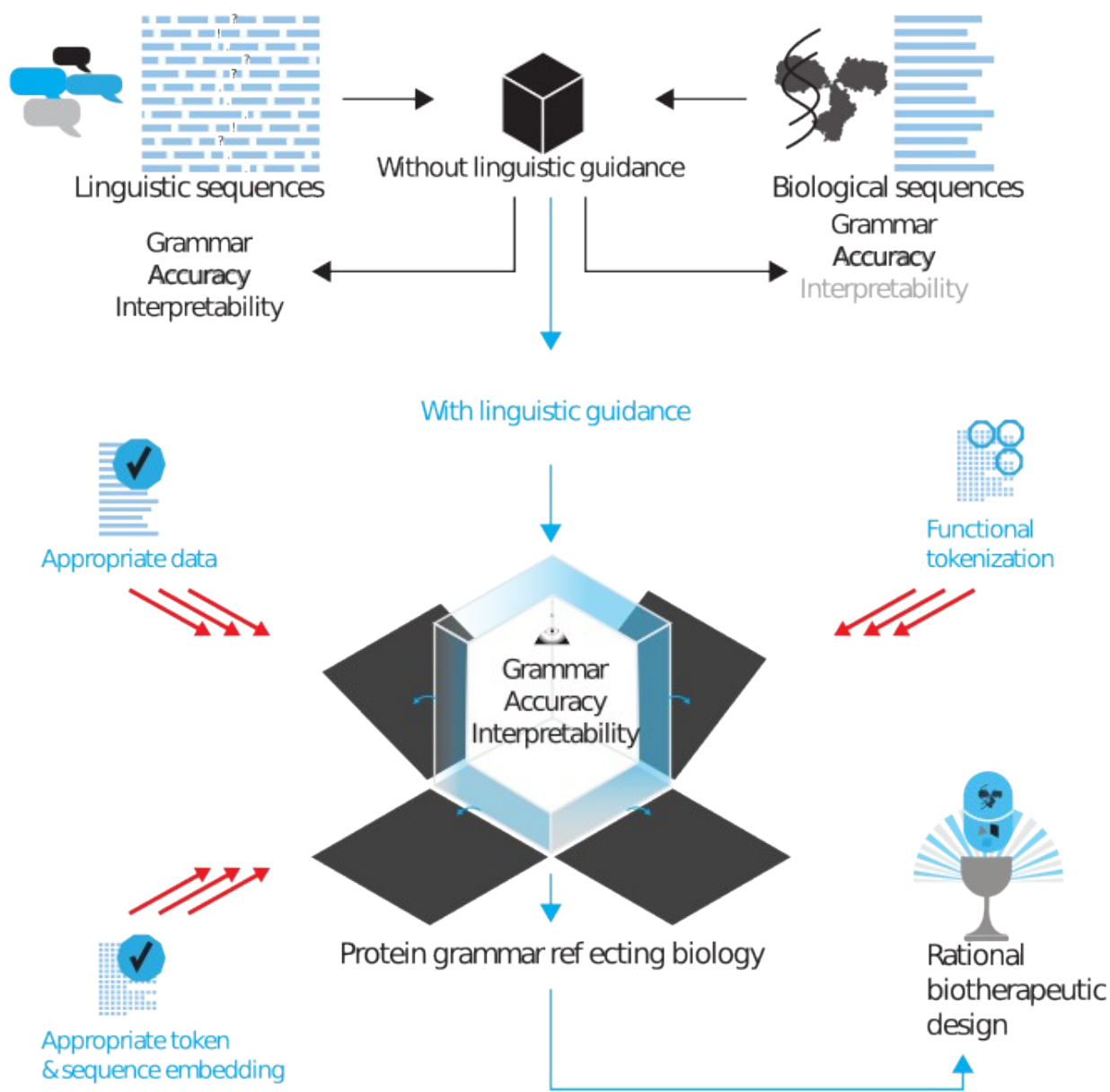

Figure 1 | **Linguistically inspired roadmap for building biologically reliable protein language models.** A direct application of LMs to protein sequences without any linguistic guidance in the design yields an opaque black-box model. While this protein model might perform with high Accuracy (defined as high performance on target task), it is unlikely to contain relevant protein Grammar (i.e., a generalization of proteins that matches biological reality, similar to natural language grammars as generalizations of linguistic sequences) and it remains low on Interpretability (i.e., a degree to which human users can understand the model and extract rules from it). In comparison, linguistic sequence models are more likely to learn a Grammar that matches independent linguistic analysis even without explicit linguistic guidance because linguistic data already contains structural indicators of basic linguistic units (e.g., punctuation, space), while protein sequence data does not. Even so, linguistic sequence models remain low on Interpretability without linguistic guidance, as domain knowledge is necessary to guide rule extraction from the model. An additional challenge that protein sequence modeling faces compared to linguistic sequence modeling is the absence of larger context beyond sequence. To remedy the disadvantages, in this Perspective we argue for linguistics-guided domain knowledge incorporation (appropriate pre-training data selection, tokenization, token and sequence embedding) into protein LMs. Namely, a thorough linguistic examination of natural language LM design can inform biologically appropriate

protein LM design and can yield interpretable, glass-box (transparent) LMs with a protein grammar that reflects biological domain knowledge. Extracting this protein grammar would facilitate rational biotherapeutics design.

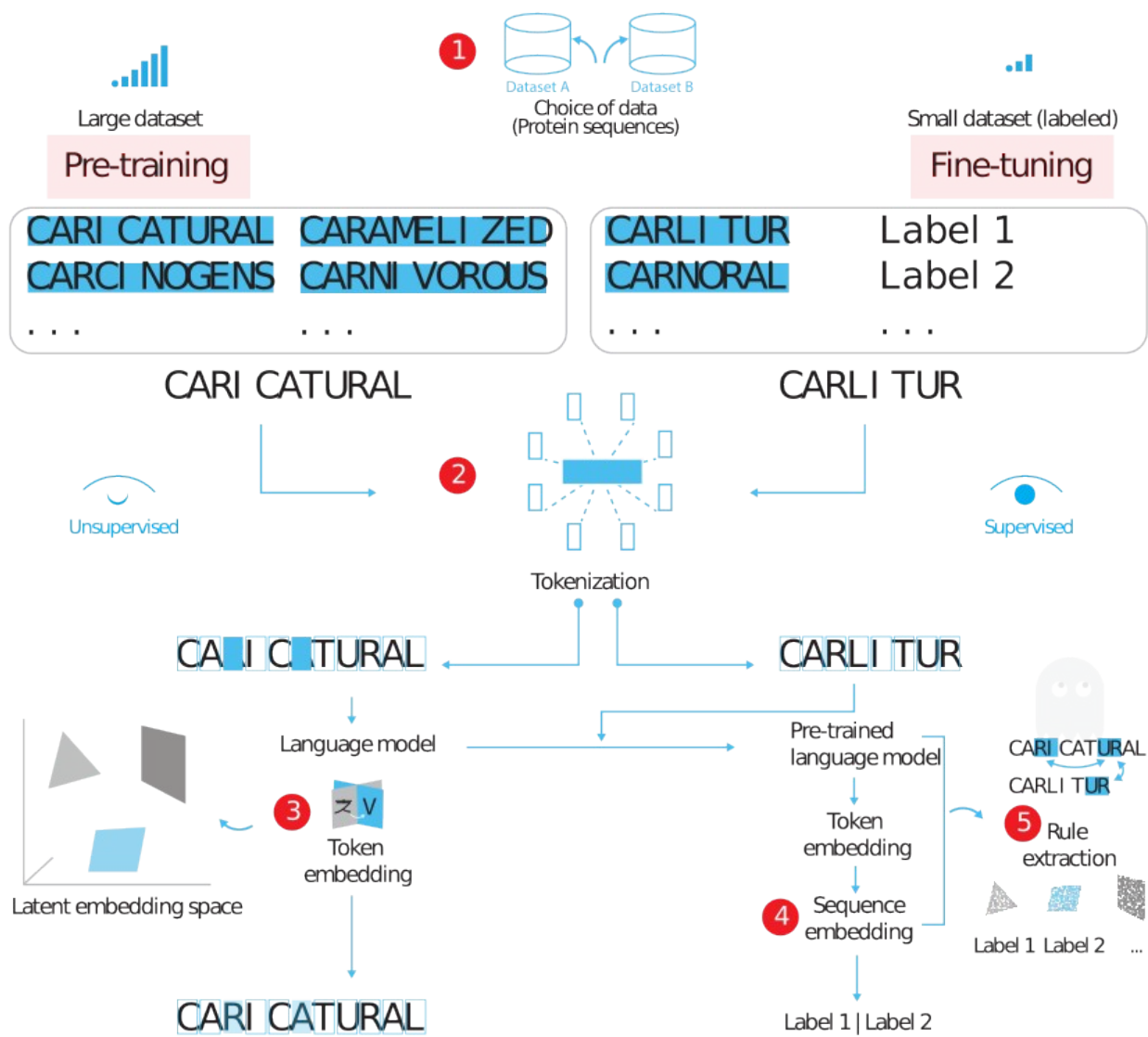

Figure 2 | **Overview of a deep language model pipeline applied to protein sequences**. An ML architecture is pre-trained in a self-supervised manner (Pre-training), independently of the task of interest on large sequencing data (1). Subsequently, the pre-trained model with added layers is trained to perform the task of interest, e.g., classification (Fine-tuning). Fine-tuning can involve tasks that require supervision and hence labeled (e.g., protein function, disease, clinical outcomes) and smaller datasets. Both steps require Tokenization (2) that segments sequences into discrete elements, usually single amino acids, due to the lack of task-informed or biologically meaningful tokens. Pre-training assigns a latent embedding to the tokens (3) that represent their contextual usage in the language. The token embedding is leveraged during fine-tuning, and the sequence embedding is calculated (4) if the fine-tuning task is a form of sequence-based prediction. Interpretation of fine-tuned models, which so far remains in its initial stages, would enable sequence-function rule discovery, such as function-associated long-range sequence dependencies (5).

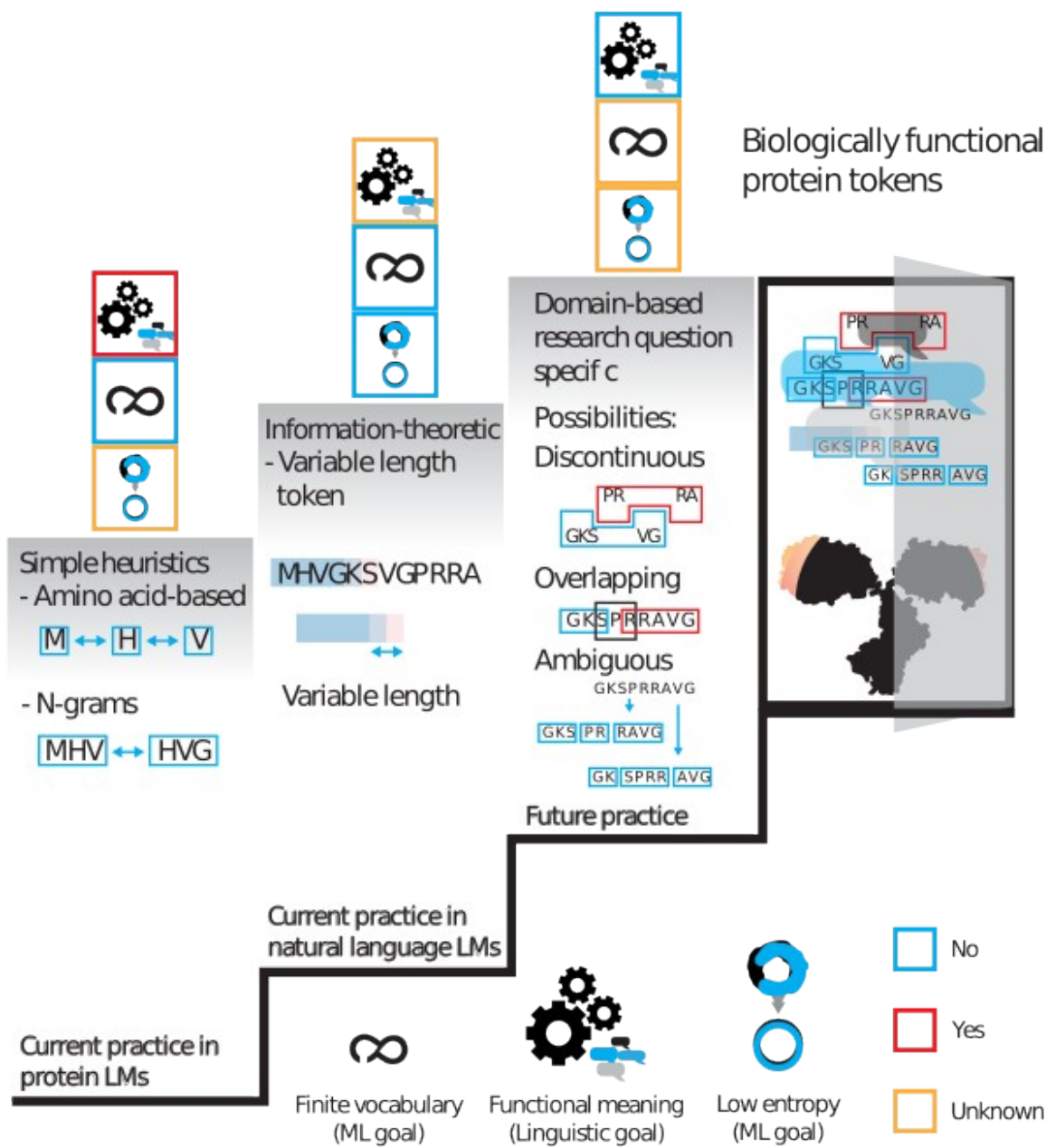

Figure 3 | **Advancing protein sequence tokenization from currently popular simple heuristics to complex methods that would generate biologically functional protein tokens akin to linguistically sound tokens in natural language.** Tokenization methods must balance three distinct goals. Linguistically sound tokens should atomically map to well-defined, abstract functional meaning. Technical constraints in ML necessitate that the generated set of possible tokens is finite and small in number (finite vocabulary), and that tokenization yields an LM with low entropy for a fixed vocabulary size (low entropy). Current practice in protein LMs is to use simple heuristics that result in tokens based on single amino acids or n-grams. While such simple heuristics yield a finite and small vocabulary, they do not map to functional meaning and it is unclear how low the generated LM entropy is. Information-theoretic tokenization methods are one step more complex, and are currently popular in natural language LMs. They also result in a finite, though larger vocabulary than simple heuristics do, and they generate low entropy LMs, but it is still unclear whether they would map to functional meaning in proteins. Finally, the most challenging and so far unimplemented method is domain-based tokenization that is specific to a research question. The tokens yielded with this method map to well-defined functional meaning, but might potentially result in an arbitrarily large, practically infinite vocabulary. The entropy of the resulting protein LM is yet unknown. Research from NLP, however, has shown that meaningful tokens lead to less redundancy,

the fraction of model entropy compared to maximum entropy[65]. Thus, if biologically defined protein tokens are information theoretically similar to linguistically meaningful natural language tokens, they would be expected to yield relatively low entropy as well. It is yet to be seen how domain-based tokens manifest, but they might be discontinuous, overlapping, and they might be ambiguous, meaning that there might be multiple possible segmentations for a given sequence. Domain-based tokenization is closest to biologically sound protein tokens akin to linguistically defined tokens in natural language.

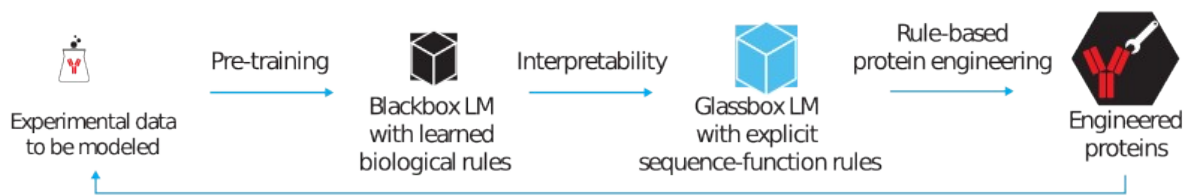

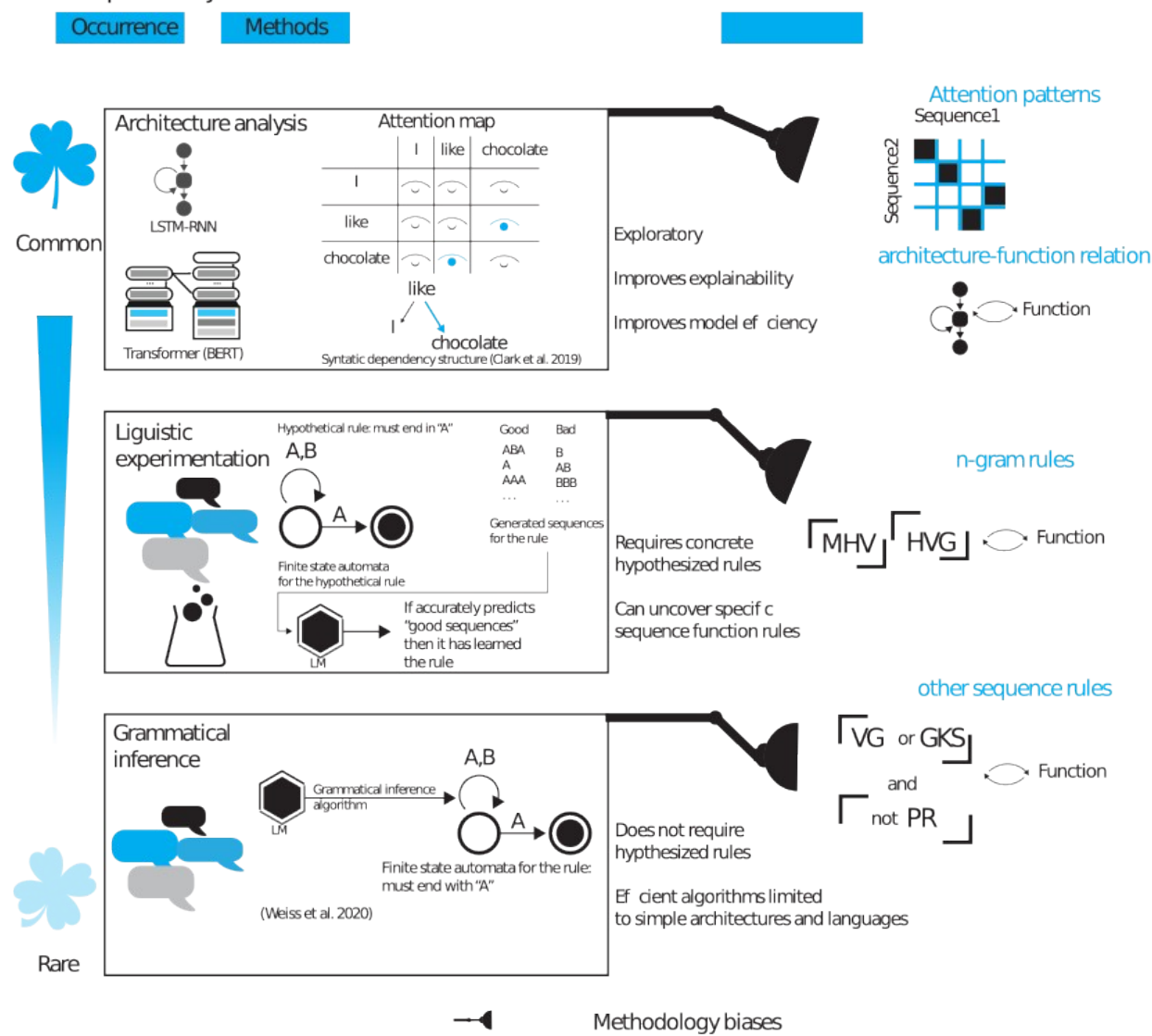

Figure 4 | **Interpretability methods for protein LMs.** (**A**) **Rule-based protein engineering workflow with protein LMs.** Black-box protein LMs with incorporated biological domain knowledge and carefully pre-trained to model the appropriate

experimental data learn biological sequence-function rules present in the pre-training data. Rule- and pattern-extraction methods can increase the explainability and interpretability of the LM, resulting in a glass-box LM with interpretable sequence-function rules. Sequence-function rules then can be leveraged for protein engineering. Engineered proteins that have been experimentally validated can potentially be added to the training data for improving protein LMs. **(B) Interpretability methods bias the nature of discoverable rules.** Different interpretability methods highlight different types of information about the architecture and the sequences. **Architecture analysis** (1) is the most commonly used method with current protein LMs for explaining black-box LMs[25,40,48,56,102], but it can only yield shallow correlational patterns that do not have a transparent and straightforward connection to biologically explanatory sequence-function rules. If protein scientists already possess ground truth that they can compare with architectural patterns, then architectural analysis can provide a better exploratory understanding of the architecture itself. However, this method cannot be reliably used for extracting new knowledge about protein sequences. A better understanding of the architecture can nevertheless be useful for improving the explainability and efficiency of the model. **Linguistics-inspired experimentation** (2)[103–108], aims to find a correlation between already known rules and the patterns learned by the LM. In this paradigm, the LM is fed with hand-crafted sequences that either follow or violate a hypothesized rule (e.g., "every sequence must end with 'A'"). This method requires that the user has a hypothesis to test, which can be challenging due to the vast number of possible rules. Studies aimed at examining the type of rules various deep neural network architectures are capable of learning [109,110] can provide a way to limit the search space for rules, though the limits might still not be sufficiently restrictive for exhaustive rule extraction. **Grammar inference** methods for deep neural networks (3), which can extract rules of a predefined power[111–113] do not require an a priori hypothesis for a concrete rule, but efficient algorithms as of now are limited to simpler architectures (e.g., RNNs) and are not yet practically suitable for more large-scale rule extraction.